\documentclass[seceq]{ptptex}

\usepackage{graphicx}
\newcommand{\Psfig}[2]{\includegraphics[width=#2]{#1}}
\usepackage{amsmath,amssymb}

\newcommand{\vc}[1]{\mbox{\boldmath $#1$}}
\newcommand{\vcs}[1]{\mbox{\footnotesize \boldmath $#1$}}
\newcommand{\nuc}[2]{\mbox{$^{#1}${#2}}}

\newcommand{\fpi}{\mbox{$f_{\pi N}$}}

\title{%        %You can use \\ for explicit line-break
Antisymmetrized Molecular Dynamics
with Coherent State Pion \\
and Its Application to Excited Spectrum of $^{12}$C
}

\author{%       %Use \scshape  for the family name
Akinori \textsc{Isshiki},
Kenichi \textsc{Naito}$^*$
and Akira \textsc{Ohnishi}
}

\inst{%         %Affiliation, neglected when [addenda] or [errata]
Division of Physics,
Graduate School of Science,
Hokkaido~University, 
Sapporo 060-0810, Japan
\\
$^*$ Meme Media Laboratory, Hokkaido~University,
Sapporo 060-8628, Japan
}

\recdate{\today}

\abst{%
We have introduced coherent state neutral pion
into Antisymmetrized Molecular Dynamics.
With the aid of coherent state technique, it becomes possible to
calculate transition matrix elements of the pion field operator
and to study excited states containing pions.
For large pion-nucleon coupling $\fpi \gtrsim 1.6$,
pions have a finite expectation value and bring large energy gain
in $^{12}$C.
We discuss two aspects of pionic effects in spectroscopy;
the $LS$ interaction like effect and the mixing of different nucleon
parity states, which would modify low energy nuclear levels.
}

\begin{document}

\maketitle

%...[%%%%%%%%%%%%%%%%%%%%%%%%%%%%%%%%%%%%%%%%%%%%%%%%%
\section{Introduction}
%%%%%%%%%%%%%%%%%%%%%%%%%%%%%%%%%%%%%%%%%%%%%%%%%%%%%%

Nuclei have been basically understood as nucleon many-body systems
in which nucleons move in a mean field
and interact via small residual interactions.
In shell models, 
each nucleon single particle wave function is assumed to have
its own orbital angular momentum, spin and parity ($lj\pi$), 
provided that the spherically symmetric mean field consists of
central and spin-orbit ($LS$) parts.
This basic picture of nuclei has been successful
in describing low-lying states of most of medium to heavy nuclei,
while there are several exceptions such as clustering states in light nuclei.

On the other hand,
the long range part ($r \gtrsim 2$ fm) of the bare nucleon-nucleon potential 
is described by the one-pion exchange potential (OPEP)
having strong tensor part, which mixes different partial waves.
This mixing is essential for the binding of deuteron,
but makes it difficult to treat exactly in many-body systems.
Thus tensor force has been usually treated in the form of
effective central and $LS$ forces in solving nuclear many-body problems
in spite of its importance in bare nuclear force,
hoping that the explicit role of the tensor force is not large in nuclei.
Actually, the first order effects of tensor force vanish in the Fermi gas state,
due to the cancellation in $(2J+1)$-weighted sum.

If pions have expectation values in nuclei or in nuclear matter,
the cancellation does not work and tensor force may play a dominant role.
There have been a lot of discussions on the possibilities of pion condensation
in nuclear matter.
For some time, it was considered that strong nucleon-$\Delta$ short range
repulsion might suppress pion condensation to emerge,
provided that
the Landau-Migdal parameter follows the "universality",
$g'_{N\Delta} = g'_{NN} \simeq 0.6$,
and that it is density-independent~\cite{Kunihiro-picond}.
Recent observations of the non-quenching in Gamow-Teller giant resonance
sum rule~\cite{Wakasa-1997} clarified that the short range repulsion
between nucleons and $\Delta$ resonances is not very strong,
$g'_{N\Delta} < 0.25$\cite{Suzuki-1999},
suggesting that the pion condensation would come in exist at least 
in high-density nuclear matter.
Also in recent {\it ab-initio} calculations of light nuclei ($A \leq 12$)
with realistic bare $NN$ potentials,
it has been shown that the OPEP contribution is dominant
in the total potential energy.\cite{Pieper:2002ne}
This result suggests that the cancellation in the Fermi gas
is not working well in actual nuclei
and that it would be necessary to consider the explicit role of pions
more seriously.

On these backgrounds, 
pion condensation and explicit role of tensor force in nuclei
has now attracting renewal interests.
In a relativistic mean field framework,
it is demonstrated that neutral pions can condensate
in the surface region of nuclei,
and that this condensation enhances the binding energy
of $jj$ closed nuclei, such as $^{12}$C~\cite{Toki2002}.
In their model,
single particle states are first prepared to have fixed $lj\pi$,
and those states having the same $j$ are mixed to gain the potential
energy from pions.
This mixture also plays a role of $LS$ like potential.\cite{Ogawa2004,Myo2005}
Since pions mix different parity (but same $j$) states,
the yrast single particle states
(the lowest energy single particle states for a given $j$,
 $s_{1/2}, p_{3/2}, d_{5/2}, f_{7/2}, g_{9/2}, h_{11/2}, i_{13/2}$)
will have the largest energy gain.
On the other hand, 
those states having smaller $j$ at around the Fermi energy will be pushed up
from the mixing with the lower energy single particle states.
It is interesting to note that the last four yrast single particle states 
are in charge of the nuclear magic numbers of 28, 50, 82 and 126.
More recently, Charge and Parity Projected Hartree-Fock (CPPHF) method 
has been developed in order to take account of
the coupling of proton and neutron single particle states
generated by OPEP.\cite{Sugimoto2004}
It has been shown that the charge projection enhances the tensor contribution
by around three times
in the case of $^4$He nuclei.

At this stage,
it would be desirable to extend the scope
of pion and tensor force study 
from the ground state and single particle states
to nuclear excited level spectroscopy with specified $J^\pi$,
which has richer information on wave functions.
Specifically, we are more interested in constructing a framework
in which explicit pionic degrees of freedom are incorporated,
rather than introducing tensor interaction,
since we believe that it is more fundamental to describe nuclear many-body
system with pions.

In this work,
we introduce coherent state pions~\cite{Amado1994}
into Antisymmetrized Molecular Dynamics (AMD)~\cite{Ono1992,Enyo1995}
and discuss the pionic effects on excited states of $^{12}$C.
Pion coherent state enables us to calculate matrix elements
of the pion operator with different states
as well as the expectation value with a given state.
In AMD, nuclear wave function is represented by the Slater determinant
of nucleon Gaussian wave packets, which is wide enough
to describe clustering states as well as shell model states.
By using the product of nucleon AMD state and pion coherent state,
we can evaluate the transition matrix element of the Hamiltonian operator
containing nucleon and pion operators.
Therefore, it becomes possible
to project the wave function to the eigen state of given $J^\pi$
and
to diagonalize the Hamiltonian matrix
consisting of wave functions with different nucleon and pion configurations.

%...]%%%%%%%%%%%%%%%%%%%%%%%%%%%%%%%%%%%%%%%%%%%%%%%%%

%...[%%%%%%%%%%%%%%%%%%%%%%%%%%%%%%%%%%%%%%%%%%%%%%%%%
\section{AMD with Coherent State Pion}
%%%%%%%%%%%%%%%%%%%%%%%%%%%%%%%%%%%%%%%%%%%%%%%%%%%%%%

The nucleon-pion basis state is assumed to be the product 
of the nucleon AMD state~\cite{Ono1992,Enyo1995}
and the pionic coherent state,\cite{Amado1994}
\begin{equation}
\label{Eq:AMD-CSpi}
 | \Psi(\vc{Z},f) \rangle
	= | \Psi_{\rm AMD}(\vc{Z}) \rangle 
	    \otimes | \Phi_\pi(f) \rangle 
\ .
\end{equation}

%%%%%%%%%%%%%%%%%%
AMD wave function is a Slater determinant
of nucleon Gaussian wave packets,
\begin{eqnarray}
 | \Psi_{\rm AMD}(\vc{Z}) \rangle 
	&=& {\cal A} \prod_i
		| \psi_{z_i} \rangle 
		| \chi^\sigma_i \chi^\tau_i\rangle
\quad (\vc{Z} = \lbrace \vc{z}_i\ ; i=1, 2, \ldots A\rbrace)\ ,\\
 \langle\vc{r}|\psi_z\rangle
	&=& \left(\frac{2\nu}{\pi}\right)^{3/4}
		\exp\left[ -\nu(\vc{r}-\vc{z}/\sqrt{\nu})^2
		+ \vc{z}^2/2 \right]
\ ,
\end{eqnarray}
where $|\chi^\sigma_i \chi^\tau_i\rangle$ represents
spin-isospin wave function.

Pion coherent state introduced by Amado et al.~\cite{Amado1994}
is represented as,
\begin{equation}
\label{Eq:PionCoherentState}
 | \Phi_\pi(f) \rangle
	= \exp \left[ \int d\vc{k} f(\vc{k}) 
		\hat{a}^\dagger(\vc{k}) \right] | 0 \rangle 
\ .
\end{equation}
By setting the commutation relation of the annihilation and creation 
operator, $\hat{a}(\vc{k})$ and $\hat{a}^\dagger(\vc{k}')$, as 
$[\hat{a}(\vc{k}), \hat{a}^\dagger(\vc{k}')] = \delta(\vc{k}-\vc{k}')$,
we can show that the above pion coherent state is an eigen state 
of the positive frequency operator $\hat{\phi}^{(+)}$,
\begin{eqnarray}
&& \hat{\phi}^{(+)}(\vc{r},t)
	= \int \frac{\hbar c\ d\vc{k}}{\sqrt{(2\pi)^3 2\omega_k/c}}
	 \hat{a}(\vc{k}) e^{i \vcs{k} \cdot \vcs{r} - i\omega_k t}
\ ,\\
&& \hat{\phi}^{(+)}(\vc{r},t) | \Phi_\pi(f) \rangle
	= \varphi_f(\vc{r},t) | \Phi_\pi(f) \rangle 
\ ,\\
&& \varphi_f(\vc{r},t)
	= \int \frac{\hbar c\ d\vc{k}}{\sqrt{(2\pi)^3 2\omega_k/c}}
	 f(\vc{k}) e^{i \vcs{k} \cdot \vcs{r} - i\omega_k t}
\ .
\end{eqnarray}
Since the bra state is an eigen state of the negative frequency operator,
$\hat{\phi}^{(-)}(\vc{r},t)$,
and the pion operator is a sum of $\hat{\phi}^{(+)}$ and $\hat{\phi}^{(-)}$,
we can easily calculate the transition matrix element of
the pion operator as follows,
\begin{equation}
\hat{\phi}(\vc{r},t) 
	= \hat{\phi}^{(+)}(\vc{r},t)
	 +  \hat{\phi}^{(-)}(\vc{r},t)
\ ,\quad
 \hat{\phi}^{(-)}(\vc{r},t)
	= \left(\hat{\phi}^{(+)}(\vc{r},t)\right)^\dagger
\ ,
\end{equation}
\begin{equation}
\langle \Phi_\pi (f) | \hat{\phi} (\vc{r},t) | \Phi_\pi (g) \rangle
= {\cal N}_\pi(\bar{f},g)
\times \left(\overline{\varphi}_f (\vc{r},t) + \varphi_g (\vc{r},t)\right)
\ ,
\end{equation}
\begin{equation}
{\cal N}_\pi(\bar{f},g)
	\equiv \langle \Phi_\pi (f) | \Phi_\pi (g) \rangle
= \exp\left[ \int d\vc{k} \bar{f}(\vc{k}) g(\vc{k}) \right]
\ .
\end{equation}

Now we consider the following Hamiltonian 
of $N$-body nucleon and pion system
containing the second quantized pion operator $\hat{\phi}$
in the axial vector $P$-wave pion-nucleon coupling,
\begin{eqnarray}
 H
	&=& \sum_{i=1}^{N} \frac{\vc{p}_i^2}{2m}
	 + \sum_{i<j} V(\vc{r}_{ij})
	+ \frac{1}{2 \hbar c } \int d\vc{r} \left[
	\vc{\nabla} \hat{\phi} (\vc{r}) \cdot \vc{\nabla} \hat{\phi} (\vc{r})
	+ \mu_\pi^2 \hat{\phi}^2(\vc{r})
		\right]
\nonumber\\
	&&+ \sum_{i=1}^{N} \frac{ \fpi }{\mu_\pi} \tau_0 
	 \left(\vc{\sigma}_i \cdot \vc{\nabla}_i\right) \hat{\phi}(\vc{r}_i)\ ,
\label{Eq:Hamiltonian}
\end{eqnarray}
where $\mu_\pi = m_\pi c/\hbar$,
and we have omitted the time dependence in the pion part.
The matrix element of this Hamiltonian is evaluated as
\begin{equation}
 {\cal H} = 
 \frac
	{ \langle \Psi(\vc{Z},f) | \hat{H} | \Psi'(\vc{Z}',g) \rangle }
 	{ \langle \Psi(\vc{Z},f) | \Psi'(\vc{Z}',g) \rangle } 
 = {\cal H}_N(\bar{\vc{Z}},\vc{Z}')
 + {\cal H}_\pi(\bar{f},g)
 + {\cal H}_{\pi N} (\bar{\vc{Z}},\vc{Z}',\bar{f},g)
\ ,
\end{equation}
\begin{equation}
\label{Eq:HamiltonianPi}
 {\cal H}_\pi(\bar{f},g)
	= \int \frac{d\vc{r}}{2\hbar c}
	\left[
	\left\{\vc{\nabla}\left(
		\overline{\varphi}_f(\vc{r})+\varphi_g(\vc{r})
		\right)\right\}^2
	+ \mu_\pi^2 \left\{
		\overline{\varphi}_f(\vc{r})+\varphi_g(\vc{r})
		\right\}^2
	\right]
\ ,
\end{equation}
\begin{equation}
\label{Eq:HamiltonianPiN}
 {\cal H}_{\pi N} (\bar{\vc{Z}},\vc{Z}',\bar{f},g)
	= \frac{ \fpi }{\mu_\pi} \int d\vc{r} 
		\vc{S}(\vc{r})
		\cdot
		\vc{\nabla}
			\left(
			 \overline{\varphi}_f(\vc{r})
			+\varphi_g(\vc{r})
			\right)
\ ,
\end{equation}
\begin{equation}
\vc{S}(\vc{r})
	=
 \frac{ \langle\Psi_{\rm AMD}(\vc{Z})|
	\ \sum_i \vc{\sigma}_i \tau_{0i} \delta(\vc{r}-\vc{r}_i)\
	|\Psi_{\rm AMD}(\vc{Z}')\rangle }
      { \langle\Psi_{\rm AMD}(\vc{Z})| \Psi_{\rm AMD}(\vc{Z}') \rangle }
\ .
\end{equation}
Here ${\cal H}_N$ is the usual AMD Hamiltonian matrix element
including $NN$ interaction.

In the actual calculation, we have expanded the pion eigen function
$\varphi(\vc{r})$ in local Gaussians, whose centers and amplitudes
are the variation parameters.
Thus we can apply the cooling equations for these pion parameters
and nucleon phase space parameters $\vc{z}_i$'s.
We have made the non-relativistic approximation
in the calculation of pionic state norm, 
\begin{equation}
{\cal N}_\pi(\bar{f},g)
\simeq \exp\left[ {2m_\pi c^2 \over \hbar^3}
	\int d\vc{r} \bar{\varphi}_f(\vc{r}) \varphi_g(\vc{r}) \right]
\ .
\end{equation}

The Hamiltonian form of Eq. (\ref{Eq:Hamiltonian})
with pion-nucleon $P$-wave interaction
is the simplest one.
In addition to coupling with charged pions,
higher dimension terms such as pion-nucleon $S$-wave interaction
coming from $\bar{N}\phi^2N$ coupling would have visible contributions
when pions have large expectation values.
Charged pions should give similar energy gains to neutral pions,
but coherent state treatment of charged pions mixes different charge
states, and this charge mixing may lead to serious problems
in the discussion of excited levels.
In order to overcome this problem, it is necessary
to perform the coupled channel calculation of different nucleon
and pion charge states or to perform isospin projection.\cite{Dote2004}
These are beyond the scope of this paper, and will be discussed elsewhere.
On the other hand, 
higher dimension terms such as the pion-nucleon $S$-wave interaction
are not expected to give large contributions in energy,
since the number of pions is around $0.1$ in $^{12}$C nuclei
in the present framework as is shown later.

%...]%%%%%%%%%%%%%%%%%%%%%%%%%%%%%%%%%%%%%%%%%%%%%%%%%

%...[%%%%%%%%%%%%%%%%%%%%%%%%%%%%%%%%%%%%%%%%%%%%%%%%%
\section{An Example of Application --- $^{12}$C Nucleus ---}
%%%%%%%%%%%%%%%%%%%%%%%%%%%%%%%%%%%%%%%%%%%%%%%%%%%%%%

We have applied AMD with coherent state pion
to the study of $^{12}$C nuclei.
In the ground state of \nuc{12}{C},
nucleons occupy the single particle states of $0s_{1/2}$ and $0p_{3/2}$,
both of which are the yrast single particle states,
then the pionic effects are expected to be large.
%
% 2_1^+
%
While the 3$\alpha$ cluster model generally describes the excited levels of
this nucleus very well,
the first excited state ($2_1^+$) is calculated to be too low.
Since the "spin-orbit" splitting of the $0p_{3/2}$ and $0p_{1/2}$
single particle states is responsible to the $0_1^+$-$2_1^+$ level spacing,
pionic effect to push up the $0p_{1/2}$ level may appear as
the increase of $E^*(2_1^+)$ in \nuc{12}{C}.
%
% 0_1^-
%
It would be also interesting to study 
un-natural parity levels such as $0^-, 1^+, 2^-, \ldots$,
whose excitation energy might decrease due to the coupling
to the natural parity nucleon state with $0^-$ pionic state.

In this paper,
we apply the simplest model of AMD with coherent state pions
as the first step.
We show the results with projection after variation (PAV);
we first construct the intrinsic state
by using the cooling variational method for the parametrized
wave function of Eq. (\ref{Eq:AMD-CSpi}),
and projection to specified $J^\pi$ has been carried out from
the prepared intrinsic state.
We find that the effects of parity projection before variation
in $^{12}$C nuclei are not large when pions are included explicitly,
while it has been found to be important
for spectroscopic studies of light nuclei without pions.\cite{Toki2002,Enyo1995}
In the cooling stage of intrinsic energy,
the imaginary part of $\varphi_f(\vc{r})$ is a redundant degree of freedom
as is clear 
from Eqs.~(\ref{Eq:HamiltonianPi}) and (\ref{Eq:HamiltonianPiN})
with $\varphi_f = \varphi_g$,
then we cannot control the imaginary part which is given randomly
in the initial state of variation.
Thus we have limited the pion eigen function $\varphi(\vc{r})$ to be real.
We find that there are many local minima especially for small $\fpi$ values,
then we have selected the lowest energy wave functions
from several candidates obtained from different random seeds.
Since we do not take account of the isospin projection
which enhances the tensor force effect by around three times,
we use larger $\fpi$ value in the range $\fpi=1 \sim \sqrt{3}$.
For finite nuclei, we use the scaled coupling constant
$\fpi^{(A)} = \sqrt{(A-1)/A} \fpi$ 
in order to include approximately the effects of the Fock (exchange)
term of OPEP, which requires quantum corrections in a field description.

As the effective nucleon-nucleon interaction, 
we start from Brink--Boeker type two range Gaussian interactions,
which approximately reproduces
the binding energies of $^4$He, $^{16}$O (28.4 and 128.9 MeV, BBO1~\cite{Okabe})
or the binding energy of $^4$He and nuclear matter saturation
(28.1 MeV and $E/A=-16.8$ MeV at $\rho = 0.165 \mbox{fm}^{-3}$, BBO2). 
Since the potential energy from pions is very large,
it would be important for nucleon-nucleon interaction
to have saturation property in order to avoid collapsing.
The interaction range is chosen to be shorter
than that of, for example, the Volkov interaction.
When we include pions, we employ the nucleon-nucleon interaction BBO$\pi$,
which is a little modified from BBO2 to reproduce the ground state energy
and the first excited state energy of $^{12}$C.
Parameters of these interactions are summarized in Table~\ref{Table:BBO}.

\begin{table}
\caption{Volkov and Brink-Boeker-Okabe interaction parameters.
Central interactions are given as the form,
$V = \sum_{i=1}^2\, v_i\, \exp(-r^2/\mu_i^2) (1-M_i - M_i P_\sigma P_\tau)$.
}\label{Table:BBO}
\begin{center}
\begin{tabular}{c|rrrrrrr}
\hline
\hline
        & $\mu_1$ (fm) & $v_1$ (MeV)  & $M_1$
        & $\mu_2$ (fm) & $v_2$ (MeV)  & $M_2$
	& $f_{\pi N}$ \\
\hline
Volkov	& 1.6  & $-83.34  $ &  0.575  & 0.82 &  144.86  & $ 0.575$  &  0 \\
BBO1\cite{Okabe}
	& 1.2  & $-253.798$ &  0.2186 & 0.6  &  924.631 & $-1.551$  &  0 \\
BBO2	& 1.2  & $-258.3  $ &  0.25   & 0.6  &  950.00  & $-1.658$  &  0 \\
BBO$\pi$& 1.2  & $-256.0  $ &  0.25   & 0.6  &  950.00  & $-1.658$  &  1.63 \\
\hline
\end{tabular}
\end{center}
\end{table}

\begin{figure}
\centerline{~\Psfig{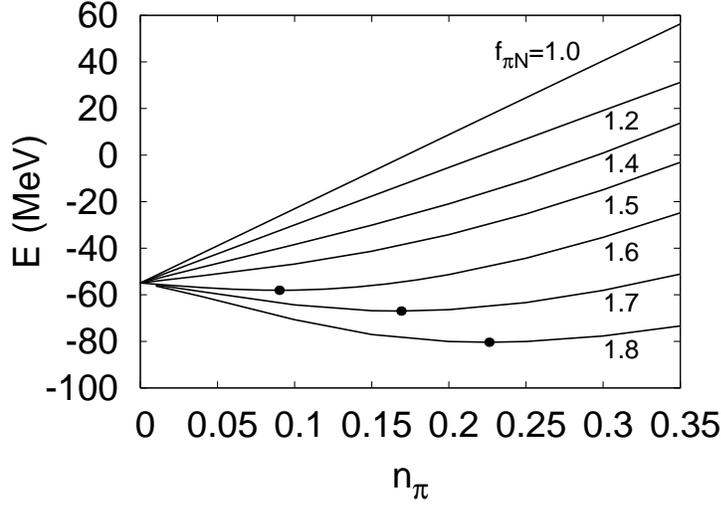}{10cm}~}
\caption{
Intrinsic energy of \nuc{12}{C} as a function of 
pion number expectation value, $n_\pi$.
Results with $\fpi = 1.0, 1.2, 1.4, 1.5, 1.6, 1.7$ and $1.8$ are shown.
Filled circles show the energy minimum for $\fpi \gtrsim 1.6$.
}
\label{Fig:Npi-Eint}
\end{figure}

\begin{figure}
\centerline{~\Psfig{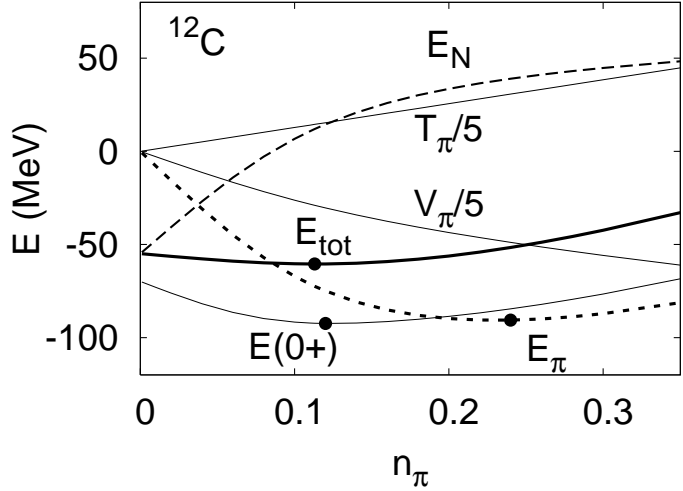}{10cm}~}
\caption{
Composition of the pionic energies for $\fpi = 1.63$.
Thick solid, long-dashed, and short-dashed lines show
total, nucleonic, and pionic energies, respectively.
Among the pionic energy,
kinetic and interaction energies
are also shown by thin solid lines.
Filled circles show the minimum points of $E_{tot}$, $E_\pi$, and $E(0^+)$.
}
\label{Fig:Npi-Vpi}
\end{figure}

In Fig.~\ref{Fig:Npi-Eint},
we show the intrinsic state energy as a function of
the pion number expectation value,
\begin{equation}
n_\pi = {
		\langle \Phi_\pi (f) |\int d\vc{k}
			\hat{a}^\dagger(\vc{k})
			\hat{a} (\vc{k}) | \Phi_\pi (f) \rangle
	\over
		\langle \Phi_\pi (f) | \Phi_\pi (f) \rangle
	}
\simeq {2m_\pi c^2 \over \hbar^3}
	\int d\vc{r} \bar{\varphi}_f(\vc{r}) \varphi_f(\vc{r})
\ .
\end{equation}
At small $\fpi$ values around one,
pure nucleon state is energetically favored.
When we increase $\fpi$,
pion-nucleon interaction gives very large binding,
and the optimal state has finite pions 
for $\fpi \gtrsim 1.6$.
The total pionic energy amounts to be around $-90$ MeV
in the case $\fpi = 1.63$ as shown in Fig.~\ref{Fig:Npi-Vpi}.
In well developed pionic states,
the nucleus loses energy in the nucleon part ${\cal H}_{NN}$
instead of gaining pion-nucleon interaction energy efficiently.
This feature is similar to the case of pion condensation
in high density nuclear matter.~\cite{Takatsuka1993}

\begin{figure}
\centerline{~\Psfig{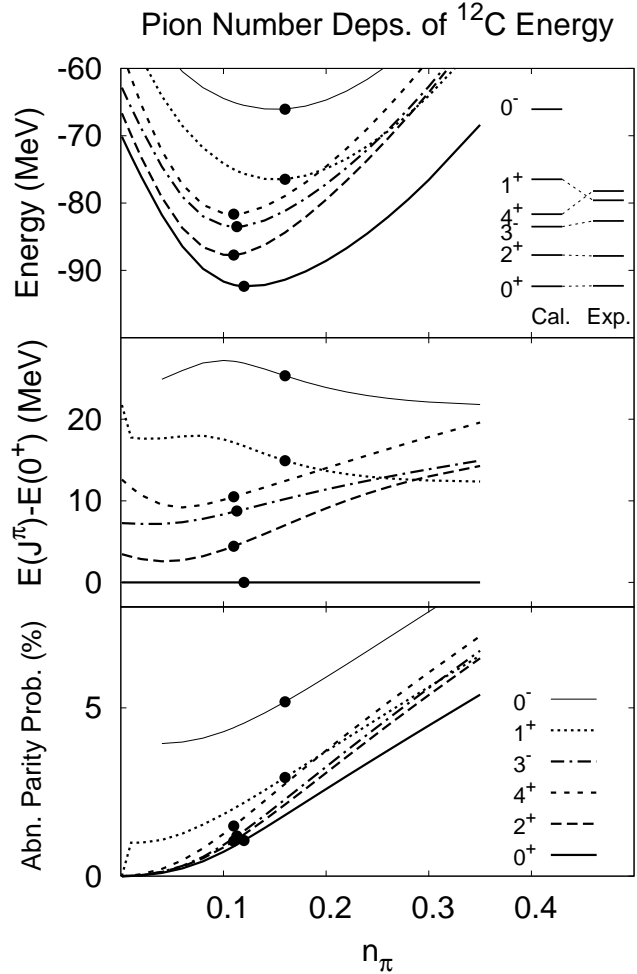}{10cm}~}
\caption{
Pion number dependence of the total energy (top),
energy difference from $0^+_1$ state (middle), 
and the nucleonic abnormal parity probability (bottom) in $^{12}$C.
Filled circles show the energy minimum points for each $J^\pi$.
}\label{Fig:Npi-ExLev}
\end{figure}

In the upper panel of Fig.~\ref{Fig:Npi-ExLev}, 
we show the results of total energy after $J^\pi$ projection
from the cooled intrinsic wave functions under $n_\pi$ constraint.
All the $J^\pi$ states have their minima at finite $n_\pi$ when we adopt
$\fpi = 1.63$.
It is interesting to find that natural parity states favor smaller $n_\pi$,
and un-natural parity states favor larger $n_\pi$.

Finite number pion is expected to act as the $LS$-like interaction
and to increase the excitation energy of $2_1^+$.
In the middle panel of Fig.~\ref{Fig:Npi-ExLev},
we show the energy difference $E(J^{\pi})-E(0^+_1)$
as functions of $n_\pi$ for $\fpi = 1.63$.
At $n_\pi = 0$ where the present model is equivalent to the normal AMD,
$2_1^+$ has small excitation energies,
which is a feature of $\alpha$ cluster models.
At around $n_\pi \simeq 0.05$, $2_1^+$ state starts to go up.
This increase of energy difference at $n_\pi > 0.1$
is not a consequence of the nuclear shrinkage,
but the result of pionic $LS$-like effect.
Actually we find that the calculated rms radius grows in the region
$n_\pi \gtrsim 0.07$.

Contrary to the positive parity rotational states,
$0_1^-$ and $1^+_1$ states go down as the pion number increases.
This is due to the coupling to the nucleonic abnormal parity states
such as,
\begin{equation}
|\Psi(0^-)\rangle
	= |\Psi_N(0^-)\rangle
	+ |\Psi_N(0^+)\rangle\otimes|\Phi_\pi(0^-)\rangle\ ,
\end{equation}
for the $0^-$ state.
In order to demonstrate this point,
we show the nucleonic abnormal parity probability $P^{{\rm Abn.}}_N$
in the bottom panel of Fig.~\ref{Fig:Npi-ExLev}.
At zero pion number $n_\pi=0$,
all the states should be purely described in nucleonic state
($P^{{\rm Abn.}}_N = 0$),
but the probability goes up to around 5 \% and 3 \%
at the projected energy minima 
for $0^-$ and $1^+$ states, respectively.
Other rotational levels are also calculated to contain
the abnormal nucleonic parity component of around 1 \%.
If these are true, it may be interesting to observe pion knock-out reaction,
which leaves the nucleus in the un-natural parity states.

%%%%%%%%%%%%

\begin{table}
\caption{
Energy components in $^{12}$C levels.
For $0^+$ state, calculated total energies are shown in the parentheses.
All the energies are shown in MeV.
}\label{Table:Levels}
\begin{center}
\begin{tabular}{c|rrrrrrr}
\hline
\hline
      &$J^\pi$ &$E^*$      &$T_N$   &$V_c$ &$V_{Coul}$ &$V_{LS}$ &$E_\pi$\\
\hline
Volkov&$0^+$   &$(-92.4)$  &$234.3$ &$-320.6$ &$8.9$   &$-15.0$ &-\\
      &$2^+$   & 4.2       &$234.5$ &$-320.6$ &$8.8$   &$-11.1$ &-\\
      &$4^+$   &12.0       &$235.4$ &$-317.5$ &$8.8$   &$ -7.2$ &-\\
      &$3^-$   &19.5       &$244.5$ &$-314.1$ &$8.7$   &$-12.0$ &-\\
\hline
BBO$\pi$
      &$0^+$   &$(-92.4)$  &$187.5$ &$-228.8$ &$7.2$   &$ 0.01$ &$-58.2$\\
      &$2^+$   & 4.6       &$190.9$ &$-233.3$ &$7.2$   &$-0.03$ &$-52.5$\\
      &$4^+$   &10.6       &$196.3$ &$-224.9$ &$7.1$   &$-0.15$ &$-60.1$\\
      &$1^+$   &17.0       &$198.1$ &$-195.3$ &$7.0$   &$ 0.25$ &$-89.6$\\
      &$0^-$   &27.0       &$204.0$ &$-183.5$ &$7.0$   &$ 0.48$ &$-89.1$\\
      &$3^-$   & 8.7       &$194.7$ &$-228.9$ &$7.2$   &$-0.04$ &$-56.5$\\
\hline
\end{tabular}
\end{center}
\end{table}

The ground state and the first $2^+$ state energy
can be reproduced in AMD without pion effects
when we adopt strong $LS$ interactions,\cite{Enyo1998}
but the wave functions in these two descriptions are very different.
In Table \ref{Table:Levels},
we compare the energy components for $^{12}$C levels
in AMD with Volkov interaction with strong $LS$ interaction ($V_{LS}=1800$ MeV)
and in the present model with moderate $LS$ interaction ($V_{LS}=900$ MeV).
The energy difference of $0^+$ and $2^+$ mainly comes from the $LS$ interaction
in the case without pions,
while the pionic energy is the main source of the energy difference
when pions are included.
In addition, it is interesting to find that the $LS$ interaction acts
in the reverse way
--- $LS$ interaction is more attractive for $0^+$ without pions,
but it weakly acts repulsively for $0^+$ with pions.

%...[%%%%%%%%%%%%%%%%%%%%%%%%%%%%%%%%%%%%%%%%%%%%%%%%%
\section{Summary}

In this paper,
we have developed a new framework to include pionic degrees of freedom
in the nuclear many-body systems, 
Antisymmetrized Molecular Dynamics (AMD) with coherent state pion.
Compared to the mean field treatment of pions,
the present model has a merit that we can evaluate
the transition matrix elements of the pion-nucleon coupling term 
containing the second quantized pion operator $\hat\phi$.
This enables us to calculate the excitation energies
for specified $J^\pi$ with explicit pion degrees of freedom,
through the parity and angular momentum projection from
the intrinsic state.
It is also interesting to find that the pion coherent state has a norm, 
and the pionic state overlap $\langle\Phi_\pi(f)|\Phi_\pi(g)\rangle$
reduces the total state overlap,
$\langle\Psi(\vc{Z},f)|\Psi(\vc{Z}',g)\rangle$.

We have applied this model to the study of \nuc{12}{C} structure.
The $LS$-like effects of pions can be seen as the increase of
the $2_1^+$ state excitation energy.
It is also suggested that explicit pionic state
$|\Psi_N(0^+)\rangle\otimes|\Phi_\pi(0^-)\rangle$
can admix to the $0^-$ state with around 5 \%
when we adopt $\fpi = 1.63$.

In order to obtain firm conclusions on the explicit pionic effects
in nuclear structure,
further theoretical and experimental developments are mandatory.
First, charged pions should be included in the framework.
Combined with the charge (isospin) projection,
it is expected to enhance the tensor effect by three times.
We have simulated this enhancement by increasing the pion-nucleon
coupling $\fpi$, but the exchange of proton and neutron
may lead to non-trivial effects which cannot be mimicked
by increasing $\fpi$.
Next, we should take care of the exchange term (Fock term)
and the zero range ($\delta$ type) part of the OPEP
in a more reliable manner than to scale the coupling constant
by $\sqrt{(A-1)/A}$ times.
Including the Landau-Migdal interaction may be an efficient way
for this problem.
Extension of the wave function is also an important direction.
Since we have assumed that the total wave function is a product
of nucleonic state and pionic coherent state,
nucleonic part of the wave function is common
to the zero and one pion states.
Thus the nucleonic part of the wave function
has to contain both of the $T=0$ and $T=1$ states,
which may lead to the underestimate of the binding energy.
Works in these directions are in progress.
%...]%%%%%%%%%%%%%%%%%%%%%%%%%%%%%%%%%%%%%%%%%%%%%%%%%

%...]%%%%%%%%%%%%%%%%%%%%%%%%%%%%%%%%%%%%%%%%%%%%%%%%%

\section*{Acknowledgements}
We would like to thank 
Prof. S. Okabe, Prof. K. Kat{\=o}, Prof. K. Ikeda, Prof. H. Toki
and Dr. T. Myo for useful discussions.
This work is supported in part by the Ministry of Education,
Science, Sports and Culture,
Grand-in-Aid for Scientific Research (C)(2), No. 15540243, 2003.

\end{document}